\documentclass[showpacs,preprintnumbers,twocolumn,
amsmath,amssymb,groupedaddress,superscriptaddress]{revtex4}

\usepackage{graphicx}
\usepackage{dcolumn}
\usepackage{bm}

\begin{document}

\title{Volume and surface contributions to the nuclear symmetry energy
within the coherent density fluctuation model}

\author{A. N. Antonov}
\affiliation{Institute for Nuclear Research and Nuclear Energy,
Bulgarian Academy of Sciences, Sofia 1784, Bulgaria}

\author{M. K. Gaidarov}
\affiliation{Institute for Nuclear Research and Nuclear Energy,
Bulgarian Academy of Sciences, Sofia 1784, Bulgaria}

\author{P. Sarriguren}
\affiliation{Instituto de Estructura de la Materia, IEM-CSIC,
Serrano 123, E-28006 Madrid, Spain}

\author{E. Moya de Guerra}
\affiliation{Grupo de F\'{i}sica Nuclear, Departamento de
F\'{i}sica At\'{o}mica, Molecular y Nuclear,\\ Facultad de
Ciencias F\'{i}sicas, Unidad Asociada UCM-CSIC(IEM), Universidad
Complutense de Madrid, E-28040 Madrid, Spain}

\begin{abstract}
The volume and surface components of the nuclear symmetry energy
(NSE) and their ratio are calculated within the coherent density
fluctuation model (CDFM). The estimations use the results of the
model for the NSE in finite nuclei based on the Brueckner
energy-density functional for nuclear matter. In addition, we
present results for the NSE and its volume and surface
contributions obtained by using the Skyrme energy-density
functional. The CDFM weight function is obtained using the proton
and neutron densities from the self-consistent HF+BCS method with
Skyrme interactions. We present and discuss the values of the
volume and surface contributions to the NSE and their ratio
obtained for the Ni, Sn, and Pb isotopic chains studying their
isotopic sensitivity. The results are compared with estimations of
other approaches which have used available experimental data on
binding energies, neutron-skin thicknesses, excitation energies to
isobaric analog states (IAS) and also with results of other
theoretical methods.
\end{abstract}


\maketitle

\section{Introduction}
One of the most exciting topics of research in nuclear physics is
currently the nuclear matter symmetry energy that essentially
characterizes the isospin-dependent part of the equation of state
of asymmetric nuclear matter (ANM)
\cite{Centelles2009,Danielewicz2002,Shetty2007,Famiano2006}. A
natural and important way to learn more about the NSE is the
transition from ANM to finite nuclei. Experimentally, the NSE is
not a directly measurable quantity and is extracted indirectly
from observables that are related to it (see, e.g., the review
\cite{Shetty2010}). A sensitive probe of the NSE is the
neutron-skin thickness of nuclei, although its precise measurement
is difficult to be done. At present, the latter is derived from
pigmy dipole resonance measurements \cite{Klimk2007}, from data on
antiprotonic atoms \cite{Centelles2009}, giant resonances, nuclear
reactions, parity-violating asymmetry
\cite{Abrahamyan2012,Moreno2009} and others. Correlations of the
neutron-skin thickness in finite nuclei with various symmetry
energy parameters are considered in Ref.~\cite{Mondal2016}. A wide
range of works (e.g.,
\cite{Li2006,Piekarewicz2009,Vidana2009,Sammarruca2009,Dan2010})
are devoted to studies of the density dependence of the symmetry
energy in uniform matter.

The symmetry energy of finite nuclei at saturation density has
been often extracted by fitting ground state masses with various
versions of the liquid-drop mass formula within the liquid-drop
models \cite{Myers66,Moller95,Pomorski2003}, and also within other
approaches, such as the Random Phase Approximation based on the
Hartree-Fock (HF) approach \cite{Carbone2010}, on effective
Lagrangians with density-dependent meson-nucleon vertex functions
\cite{Vretenar2003}, energy density functional (EDF) of Skyrme
force \cite{Chen2005,Yoshida2006,Chen2010}, relativistic
nucleon-nucleon (NN) interactions \cite{Lee98,Agrawal2010} and
others. In our previous works \cite{Gaidarov2011,Gaidarov2012} the
symmetry energy has been studied in a wide range of spherical and
deformed nuclei, correspondingly, on the basis of the Brueckner
EDF \cite{Brueckner68,Brueckner69}. In these works the transition
from the properties of nuclear matter to those of finite nuclei
has been made using the coherent density fluctuation model
\cite{Ant80,AHP}. The latter is a natural extension of the
Fermi-gas model based on the generator coordinate method
\cite{AHP,Grif57} and includes long-range NN correlations of
collective type. The numerous applications of the CDFM to analyses
of characteristics of the nuclear structure and reactions can be
seen, e.g. in Refs.~\cite{Gaidarov2011,AHP}. In
\cite{Gaidarov2011} the study of the correlation between the
thickness of the neutron skin in finite nuclei and the NSE for the
isotopic chains of even-even Ni ($A=74-84$), Sn ($A=124-152$) and
Pb ($A=206-214$) nuclei, as well as the neutron pressure and the
asymmetric compressibility for these nuclei have been presented.
The calculations have been based on the deformed self-consistent
mean-field HF+BCS method \cite{vautherin,Guerra91}, using the CDFM
and the Brueckner EDF. The same approaches have been used in
Ref.~\cite{Gaidarov2012} for the calculations of the mentioned
quantities of deformed neutron-rich even-even nuclei, such as Kr
($A=82-120$) and Sm ($A=140-156$) isotopes.

In 1947 Feenberg \cite{Feenberg47} pointed out that the surface
energy should contain a symmetry energy contribution as a
consequence of the failure of the nuclear saturation at the edge
of the nucleus and that the volume saturation energy also has a
symmetry energy term. Cameron in 1957 \cite{Cameron57} (see also
Bethe \cite{Bethe71}) suggested a revised mass formula in which
the volume energy was expressed as a sum of two contributions, the
volume saturation energy proportional to the mass number $A$ and a
volume symmetry energy assumed proportional to $(A-2Z)^{2}/A$. In
1958 Green \cite{Green58} estimated the values of the volume and
surface components of the corresponding contributions to the
symmetry energy. Myers and Swiatecki in 1966 \cite{Myers66}
admitted that the ratio between the mentioned coefficients must be
equal to the ratio between the surface and volume coefficients of
the corresponding components of the mass formula. In Ref.~\cite
{Warda2010} Warda {\it et al.} studied the bulk and the surface
nature of the formation of the neutron skin in the isotopic chains
of Sn and Pb, a concept that can be applied when analyzing the
experimental data. In Ref.~\cite{Centelles2010} the same authors
showed the role of the stiffness of the NSE on the origin of the
neutron-skin thickness in $^{208}$Pb, the latter being decomposed
into bulk and surface components. In Ref.~\cite{Dan2003} it has
been demonstrated by Danielewicz that the ratio of the volume to
surface symmetry energy is closely related to the neutron-skin
thickness (see also
Refs.~\cite{Dan2004,Dan2014,Dan2009,Dan2011,Tsang2012,Tsang2009,Ono2004,Dan2006}).
Discussions on the correlation between the bulk and surface
symmetry energy are given also e.g., in
Refs.~\cite{Steiner2005,Danielewicz,Diep2007,Kolomietz,Nikolov2011}.
It has been shown in \cite{Lee2010} by Lee and Mekjian by
calculations of the thermal nuclear properties that the surface
symmetry-energy term is more sensitive to the temperature than the
volume energy term. In Ref.~\cite{Agrawal2014} Agrawal {\it et
al.} pointed out that contrary to the case of the infinite nuclear
matter, a substantial change in the symmetry energy coefficients
is observed for finite nuclei with temperature.

In the present work we investigate the volume and surface
contributions to the NSE within the CDFM. We use our results for
NSE obtained using Brueckner EDF in
Refs.~\cite{Gaidarov2011,Gaidarov2012,Gaidarov2014}, as well as
the considerations of this subject mentioned above (e.g.,
\cite{Feenberg47,Cameron57,Bethe71,Green58,Warda2010,Centelles2010,Dan2003,Dan2004,Dan2014,
Dan2009,Dan2011,Tsang2012,Tsang2009,Ono2004,Dan2006,Steiner2005,Danielewicz,Diep2007,Kolomietz,Nikolov2011}).
The present calculations are performed using both Brueckner and,
in addition, Skyrme energy-density functionals. The results are
compared with those of other theoretical methods and with
corresponding experimental data obtained from analyses of
different nuclear quantities, such as binding energies,
neutron-skin thicknesses, excitation energies to IAS and others.

The structure of this paper is the following. In
Sec.~\ref{sec:relationships} we present the main relationships for
the NSE and its volume and surface components that we use in our
study. Section~\ref{sec:results} contains the CDFM formalism that
provides a way to calculate the mentioned quantities. There we
also present the numerical results and discussions. The main
conclusions of the study are given in
Section~\ref{sec:conclusions}.

\section{Relationships concerning the volume and surface contributions to nuclear symmetry energy}
\label{sec:relationships}

The mass formula can be written in the form (e.g.,
Ref.~\cite{Bethe71}):
\begin {eqnarray}
E&=&-c_{1}A+c_{2}A^{2/3}+c_{3}^{\prime}\frac{(N-Z)^{2}}{A} \nonumber \\
&+& \text{Coulomb term} + \text{shell corrections}
\label{eq:1}
\end{eqnarray}
The first three terms in the right-hand side of Eq.~(\ref{eq:1})
correspond to the volume, surface and symmetry components of the
energy. As mentioned in the Introduction, the symmetry energy has
volume and surface contributions. Then, the third term in
Eq.~(\ref{eq:1}) has to be replaced by (see, e.g.,
\cite{Bethe71}):
\begin {equation}
\frac{(N-Z)^{2}}{A}(c_{3}-c_{4}A^{-1/3}).
\label{eq:2}
\end{equation}
Estimations of the coefficients $c_{3}$ and $c_{4}$ have been
given in Ref.~\cite{Green58}, but due to the substantial shell
corrections there remained problems to obtain their values. In
Ref.~\cite{Myers66} it has been admitted that the ratio
$c_{4}/c_{3}$ can be taken to be equal to the ratio $c_{2}/c_{1}$
of the coefficients of the surface to volume components of the
energy (see also \cite{Bethe71}):
\begin {equation}
\frac{c_{4}}{c_{3}}=\frac{c_{2}}{c_{1}}=\chi.
\label{eq:3}
\end{equation}
In the work of Myers and Swiatecki \cite{Myers66} the value of
$\chi$ is estimated to be 1.1838, while it is given to be 1.14 by
Bethe in Ref.~\cite{Bethe71}.

The expression for the energy per particle has the form:
\begin {eqnarray}
\bar{E}&=&\frac{E}{A}=-c_{1}+c_{2}\frac{1}{A^{1/3}}+c_{3}^{\prime}\left ( \frac{N-Z}{A} \right )^{2} \nonumber \\
&+& \frac{1}{A}[\text{Coulomb term} + \text{shell corrections}].
\label{eq:4}
\end{eqnarray}
By definition the NSE coefficient is
\begin{equation}
s=\frac{1}{2}\left. \frac{\partial^{2}\bar{E}}{\partial\alpha^{2}}
\right |_{\alpha=0},
\label{eq:5}
\end{equation}
where
\begin{equation}
\alpha\equiv \frac{N-Z}{A}.
\label{eq:6}
\end{equation}
Then it follows from Eqs.~(\ref{eq:5}), (\ref{eq:2}) and
(\ref{eq:3}) that
\begin{equation}
s=c_{3}^{\prime}=c_{3}-\frac{c_{4}}{A^{1/3}}=c_{3}\left
(1-\frac{\chi}{A^{1/3}}\right ),
\label{eq:7}
\end{equation}
and
\begin{equation}
c_{3}=\frac{s}{1-\frac{\chi}{A^{1/3}}},  \;\;\;\;\;
c_{4}=\chi\left(\frac{s}{1-\frac{\chi}{A^{1/3}}}\right ).
\label{eq:8}
\end{equation}

In modern times Danielewicz {\it et al.} (see, e.g.,
\cite{Dan2003,Dan2004,Dan2006,Danielewicz} and references therein)
proposed the following expression for the symmetry energy:
\begin {equation}
E_{sym}=\frac{a_{a}(A)}{A}(N-Z)^{2},
\label{eq:9}
\end{equation}
where the $A$-dependent coefficient $a_{a}(A)$ is expressed by
means of the volume ($a^{V}_{A}$) and surface ($a^{S}_{A}$)
coefficients in the form:
\begin{equation}
a_{a}(A)=\frac{a_{A}^{V}}{\left
[1+A^{-1/3}\frac{a_{A}^{V}}{a_{A}^{S}}\right ]},
\label{eq:10}
\end{equation}
that is also rewritten as
\begin {equation}
\frac{1}{a_{a}(A)}=\frac{1}{a_{A}^{V}}+\frac{A^{-1/3}}{a_{A}^{S}}.
\label{eq:11}
\end{equation}
As expected, the expressions [Eqs.~(\ref{eq:9}-\ref{eq:11})] are
related to that from the earlier works [Eq.~(\ref{eq:2}) and also
see Eq.~(\ref{eq:22}) in the next subsection \ref{sec:cdfm}].

Here we would like to mention that Eqs.~(\ref{eq:10}) and
(\ref{eq:2}) are used in Ref.~\cite{Agrawal2014} as "definition I"
\cite{Myers69,Dan2003,Danielewicz} and "definition II"
\cite{Lipparini82,Jiang2012,Reinhard2006}, respectively (see
Eqs.~(37) and (38) in Ref.~\cite{Agrawal2014}).

An important result that expresses the ratio of the volume to the
surface energy coefficients by means of the shape of the symmetry
energy dependence on density $s(\rho)$ is given (in the local
density approximation to the symmetry energy), e.g. in
Refs.~\cite{Dan2006,Diep2007} (see also Ref.~\cite{Dan2003}):
\begin {equation}
\frac{a_{A}^{V}}{a_{A}^{S}}=\frac{3}{r_{0}}\int dr
\frac{\rho(r)}{\rho_{0}} \left
\{\frac{s(\rho_{0})}{s[\rho(r)]}-1\right \}.
\label{eq:12}
\end{equation}
In Eq.~(\ref{eq:12}) $\rho(r)$ is the half-infinite nuclear matter
density, $\rho_{0}$ is the nuclear matter equilibrium density, and
$r_{0}$ is the radius of the nuclear volume per nucleon that can
be obtained from
\begin {equation}
\frac{4\pi r_{0}^{3}}{3}=\frac{1}{\rho_{0}}.
\label{eq:13}
\end{equation}
For density-independent symmetry energy $s(\rho)=s(\rho_{0})=
a^{V}_{A}$ and, then it follows from Eq.~(\ref{eq:12}) that the
ratio $a^{V}_{A}/a^{S}_{A}=0$ \cite{Dan2006}. The density
$\rho(r)$ in Eq.~(\ref{eq:12}) is uniform in two Cartesian
directions and generally nonuniform in the third, usually chosen
to be $z$ \cite{Danielewicz}. The integral in Eq.~(\ref{eq:12}) is
across the nuclear surface involving the shape of the density
dependence \cite{Dan2004}. In the Danielewicz's approximation only
the symmetry energy of a finite nucleus $a_{a}(A)$ has a mass
dependence, while $a^{V}_{A}$, $a^{S}_{A}$, and their ratio
$a^{V}_{A}/a^{S}_{A}$ are $A$-independent quantities. The values
of $a^{V}_{A}$ and $a^{S}_{A}$ differ for various Skyrme
interactions in wide intervals (see Table I of
Ref.~\cite{Danielewicz}). At the same time, as shown in
\cite{Dan2003}, a combination of empirical data on skin sizes and
masses of nuclei constrains the volume symmetry parameter to
$27\leq a^{V}_{A}\leq 31$ MeV and the ratio $a^{V}_{A}/a^{S}_{A}$
to $2.0 \leq a^{V}_{A}/a^{S}_{A} \leq 2.8$.

In the next Section~\ref{sec:results} we use the relationships
from this section in order to consider the volume and surface
components of the NSE and their ratio within the CDFM.

\section{The CDFM. Results of calculations of NSE and its volume and
surface contributions}
\label{sec:results}
\subsection{The CDFM scheme to calculate the NSE}
The CDFM has been introduced and developed in
Refs.~\cite{Ant80,AHP} (and references therein). In the model the
one-body density matrix $\rho({\bf r},{\bf r^{\prime}})$ of the
nucleus is written as a coherent superposition of the one-body
density matrices $\rho_{x}({\bf r},{\bf r^{\prime}})$ for
spherical "pieces" of nuclear matter (so called "fluctons") with
densities $\rho_{x}({\bf r})=\rho_{0}(x)\Theta(x-|{\bf r}|)$,
$\rho_{0}(x)=3A/4\pi x^{3}$:
\begin{equation}
\rho({\bf r},{\bf r^{\prime}})=\int_{0}^{\infty}dx |{\cal
F}(x)|^{2} \rho_{x}({\bf r},{\bf r^{\prime}})
\label{eq:14}
\end{equation}
with
\begin{eqnarray}
\rho_{x}({\bf r},{\bf r^{\prime}})&=&3\rho_{0}(x)
\frac{j_{1}(k_{F}(x)|{\bf r}-{\bf r^{\prime}}|)}{(k_{F}(x)|{\bf
r}-{\bf r^{\prime}}|)}\nonumber \\  & \times & \Theta \left
(x-\frac{|{\bf r}+{\bf r^{\prime}}|}{2}\right ),
\label{eq:15}
\end{eqnarray}
where $j_{1}$ is the first-order spherical Bessel function and
\begin{equation}
k_{F}(x)=\left(\frac{3\pi^{2}}{2}\rho_{0}(x)\right )^{1/3}\equiv
\frac{\beta}{x}
\label{eq:16}
\end{equation}
with
\begin{equation}
\beta=\left(\frac{9\pi A}{8}\right )^{1/3}\simeq 1.52A^{1/3}
\label{eq:17}
\end{equation}
is the Fermi momentum of the nucleons in the flucton with a radius
$x$. It follows from Eqs.~(\ref{eq:14}) and (\ref{eq:15}) that the
density distribution in the CDFM has the form:
\begin{equation}
\rho({\bf r})=\int_{0}^{\infty}dx|{\cal F}(x)|^{2}\rho_{0}(x)\Theta(x-|{\bf r}|).
\label{eq:17a}
\end{equation}
The weight function $|{\cal F}(x)|^{2}$ in Eq.~(\ref{eq:14}) can
be expressed using Eq.~(\ref{eq:17a}) by the density distribution
$\rho(r)$ and in the case of the monotonically decreasing local
density ($d\rho/dr\leq 0$) can be obtained using a known density
(from experiments or from theoretical models) for a given nucleus:
\begin{equation}
|{\cal F}(x)|^{2}=-\frac{1}{\rho_{0}(x)} \left.
\frac{d\rho(r)}{dr}\right |_{r=x}
\label{eq:18}
\end{equation}
with the normalization $\int_{0}^{\infty}dx |{\cal F}(x)|^{2}=1$.

The main assumption of the CDFM is that properties of finite
nuclei can be calculated by expressions (obtained by using some
approximations) that contain the corresponding quantities for
nuclear matter folded with the weight function $|{\cal
F}(x)|^{2}$. Then in the CDFM, the symmetry energy $s$ for finite
nuclei is obtained to be infinite superposition of the
corresponding ANM symmetry energy weighted by $|{\cal F}(x)|^{2}$:
\begin{equation}
s=\int_{0}^{\infty}dx|{\cal F}(x)|^{2}s^{ANM}(x).
\label{eq:19}
\end{equation}
The ANM quantity $s^{ANM}(x)$ has to be  determined within a
chosen method for description of these characteristics. In our
previous works \cite{Gaidarov2011,Gaidarov2012,Gaidarov2014} we
have used for the matrix element $V(x)$ of the nuclear
Hamiltonian, as an example, the corresponding ANM energy from the
energy density functional of Brueckner {\it et al.}
\cite{Brueckner68,Brueckner69}. The corresponding expression for
$s^{ANM}(x)$ within this method can be found in
Refs.~\cite{Gaidarov2011,Gaidarov2012,Gaidarov2014}. In these
works we have calculated $s$ using Eq.~(\ref{eq:19}). In order to
calculate the weight function $|{\cal F}(x)|^{2}$ from
Eq.~(\ref{eq:18}) we used the calculated proton and neutron
density distributions obtained from the self-consistent HF+BCS
method from Ref.~\cite{Guerra91} (see also
Ref.~\cite{Sarriguren2007}) with density-dependent Skyrme
interactions \cite{vautherin} and pairing correlations. In the
method the pairing between like nucleons is included by solving
the BCS equations at each iteration with a fixed pairing strength
that reproduces the odd-even experimental mass differences
\cite{Audi2003}. The numerical results for $s$ in spherical Ni,
Sn, and Pb isotopic chains are given in Ref.~\cite{Gaidarov2011},
while for deformed exotic neutron-rich even-even Kr and Sm
isotopes are presented in Ref.~\cite{Gaidarov2012}. The density
dependence of the NSE for neutron-rich and neutron-deficient Mg
isotopes with $A$=20--36 is studied in Ref.~\cite{Gaidarov2014}.

In the end of this subsection we have to note, in order to avoid
any misunderstanding, that the symmetry energy $s$ (as well as the
related quantities $a_{A}^{V}$, $a_{A}^{S}$, and their ratio that
are calculated in what follows in our work) are obtained within
the CDFM using firstly the energy-density functional of Brueckner
for the symmetry energy in infinite nuclear matter $s^{ANM}$ in
Eq.~(\ref{eq:19}), while the weight function $|{\cal F}(x)|^{2}$
is obtained using Eq.~(\ref{eq:18}) by means of the density
distribution calculated within the Skyrme HF+BCS method. Second,
we calculate in the CDFM $s$, $a^{V}_{A}$, $a^{S}_{A}$, and
$a^{V}_{A}/a^{S}_{A}$ using as an additional example the Skyrme
energy-density functional. In this case there is a
self-consistency between the way to obtain $|{\cal F}(x)|^{2}$ in
the Skyrme HF+BCS method and the use of the Skyrme EDF to obtain
NSE and its components.

\subsection{Calculations of the volume and surface contributions to
the NSE and their ratio within the CDFM}
\label{sec:cdfm}

In the beginning of this subsection we show, first, that the
expressions containing the coefficient $a_{a}(A)$
[Eqs.~(\ref{eq:9}) and (\ref{eq:10})] (e.g., from
\cite{Dan2003,Dan2004,Dan2006,Danielewicz,Diep2007}), as expected,
can be represented approximately in the form of Eq.~(\ref{eq:2}):
\begin{equation}
a_{a}(A)=\frac{a_{A}^{V}}{\left
[1+A^{-1/3}\frac{a_{A}^{V}}{a_{A}^{S}}\right ]}\simeq
c_{3}-\frac{c_{4}}{A^{1/3}}
\label{eq:22}
\end{equation}
that corresponds to Eq.~({\ref{eq:7}), if  $c_{3}=a^{V}_{A}$ and
$c_{4}=(a^{V}_{A})^{2}/a^{S}_{A}$. Eq.~(\ref{eq:22}) is obtained
for large $A$ (e.g., at least for $A\geq 27$).

In the present paper we develop, using as a base the Danielewicz's
model (and specifically Eq.~(\ref{eq:12})), another approach to
calculate the ratio $a^{V}_{A}/a^{S}_{A}$, as well as $a^{V}_{A}$
and $a^{S}_{A}$ within the CDFM. Our motivation is that numerous
analyses of the volume and surface components of the NSE using a
wide range of data on the binding energies, neutron-skin
thicknesses and excitation energies to IAS give estimations
(presented later in our paper) of these quantities as functions of
the mass number $A$ (e.g.,
Refs.~\cite{Moller95,Warda2010,Centelles2010,Steiner2005,Diep2007,Groote76,Koura2005})
that change in some intervals for different regions of nuclei. For
instance, the reported values of $a^{V}_{A}$ and $a^{S}_{A}$ are
consistent with each other in a wide mass region ($30\leq A\leq
240$). In the CDFM we take nuclear matter values of the parameters
to deduce their values in finite nuclei (using the
self-consistently calculated nuclear density) which become
dependent on the considered nucleus. For this purpose, we start
from Eq.~(\ref{eq:12}) but in it we replace the density $\rho(r)$
for the half-infinite nuclear matter in the integrand by the
density distribution of finite nucleus. Later, using
Eq.~(\ref{eq:17a}) we obtain approximately an expression that
allows us to calculate the ratio $a^{V}_{A}/a^{S}_{A}$. It has the
form:
\begin{equation}
\frac{a_{A}^{V}}{a_{A}^{S}}=\frac{3}{r_{0}\rho_{0}}\int_{0}^{\infty}dx
|{\cal F}(x)|^{2} x \rho_{0}(x)\left
\{\frac{s(\rho_{0})}{s[\rho_{0}(x)]}-1\right \} .
\label{eq:23}
\end{equation}
The approximations made in the CDFM lead to one-dimensional
integral over $x$, the latter being the radius of the "flucton"
that is perpendicular to the nuclear surface. Here we would like
to emphasize that in contrast to Eq.~(\ref{eq:12}), in
Eq.~(\ref{eq:23}) we use the finite nuclei densities to calculate
the weight function $|{\cal F}(x)|^{2}$. In this way, the integral
in Eq.~(\ref{eq:23}) contains shell effects (different from the
Friedel oscillations \cite{Ayachi87,Sandulescu2005} present in any
quantal calculations of semi-infinite nuclear matter) and
curvature contributions. Thus, a caution is necessary when
considering the role of these effects on the key quantity
$a^{V}_{A}/a^{S}_{A}$ ratio. The procedure to go from $A$=infinite
to finite $A$ that we follow to go from Eq.~(\ref{eq:12}) to
Eq.~(\ref{eq:23}) is the same that we have followed for other
nuclear properties within the CDFM, so there is no conceptual
inconsistency.

The purpose of our approach is to use the CDFM not only to
calculate the NSE $s$, but also the ratio $a^{V}_{A}/a^{S}_{A}$
and $a^{V}_{A}$ and $a^{S}_{A}$ separately. Thus, the calculations
of these quantities in one and the same model leads to a
self-consistency. As will be shown later in the paper, in the CDFM
the dependence of $a^{V}_{A}$ and $a^{S}_{A}$ on $A$ turns out to
be weak. The differences within a given isotopic chain are
approximately between 0.5 MeV and 1.5 MeV. They are narrower than
the differences in the Danielewicz's approach using in the
calculations different Skyrme forces (e.g., Table I of
Ref.~\cite{Danielewicz}). We note that our method is different
from that of Danielewicz. Starting from Eq.~(\ref{eq:12}), the
approximation of the CDFM enables us to use not the half-infinite
nuclear matter density but densities of finite nuclei. The results
turn out to be consistent with the large amount of empirical data,
as will be shown below. The spirit of our approach is in some
sense opposite to what it was  done in the past. For instance, in
the LDM mass formula one takes empirical mass values of finite
nuclei to extract "$A$-independent" values of the parameters, some
of which are afterwards extrapolated to nuclear matter energy
density functionals, like e.g., in the Brueckner EDF. The
Danielewicz's formalism is on these lines, but parametrizing
surface effects through half-infinite nuclear matter. In
Eq.~(\ref{eq:23}) $s(\rho_{0})=s^{ANM}(\rho_{0})$ and the quantity
$s[\rho_{0}(x)]=s^{ANM}[\rho_0(x)]$ is the NSE within the chosen
approach for the EDF. From Eqs.~(\ref{eq:18}) and (\ref{eq:19}) we
obtain the CDFM value for the NSE
\begin {equation}
s\equiv a_{a}(A).
\label{eq:24}
\end{equation}
Let denote by
\begin {equation}
\kappa \equiv \frac{a_{A}^{V}}{a_{A}^{S}}
\label{eq:25}
\end{equation}
that can be calculated using Eq.~(\ref{eq:23}). Then it follows
from Eq.~(\ref{eq:10}):
\begin{equation}
s=\frac{a_{A}^{V}}{1+A^{-1/3}\kappa }.
\label{eq:26}
\end{equation}
Finally, as a next step we obtain from
Eqs.~(\ref{eq:24})-(\ref{eq:26}) (having calculated within the
CDFM the values of $s$ and $\kappa$) the expressions from which we
can estimate the values of $a^{V}_{A}$ and $a^{S}_{A}$ separately:
\begin{equation}
a_{A}^{V}=s(1+A^{-1/3}\kappa),
\label{eq:27}
\end{equation}
\begin{equation}
a_{A}^{S}=\frac{s}{\kappa }(1+A^{-1/3}\kappa).
\label{eq:28}
\end{equation}

In our work we use two energy-density functionals. The first one
is the Brueckner EDF \cite{Brueckner68,Brueckner69}. It was used
in our previous works
\cite{Gaidarov2011,Gaidarov2012,Gaidarov2014} to calculate the NSE
$s$ using Eq.~(\ref{eq:19}). In this approach the value of the
equilibrium nuclear matter density is $\rho_{0}=0.204$ fm$^{-3}$,
$r_{0}=1.054$ fm [obtained from Eq.~(\ref{eq:13})], and the
symmetry energy at equilibrium nuclear matter density
$s(\rho_{0})$ in Eq.~(\ref{eq:23}) is equal to 35.07 MeV. In the
case of the Brueckner EDF the results of the calculations using
Eq.~(\ref{eq:23}) of the ratio $\kappa$ as a function of the mass
number $A$ for the isotopic chains of Ni, Sn, and Pb with
different forces (SLy4, SGII, and Sk3) are given in
Figs.~\ref{fig1}-\ref{fig3}, respectively. By means of
Eqs.~(\ref{eq:27}) and (\ref{eq:28}) and the values of the NSE
obtained in our works \cite{Gaidarov2011,Gaidarov2012}, we
calculated the coefficients $a^{V}_{A}$ and $a^{S}_{A}$. Their
values as functions of $A$ for the same isotopic chains are
presented in Figs.~\ref{fig4}-\ref{fig6}, respectively.

\begin{figure}
\centering
\includegraphics[width=78mm]{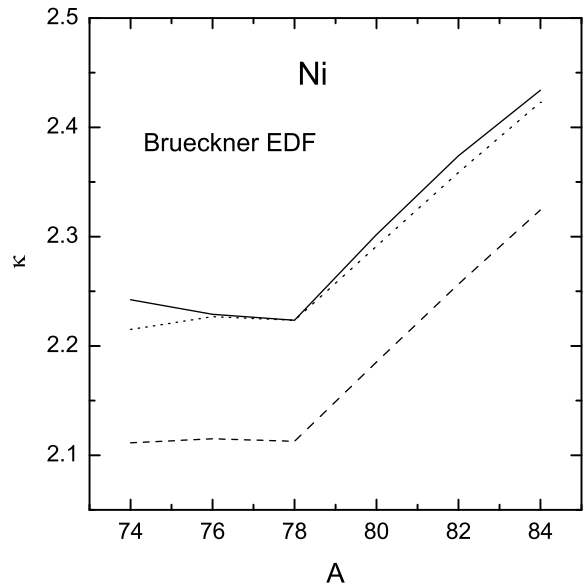}
\caption[]{The ratio $\kappa=a^{V}_{A}/a^{S}_{A}$ as a function of
$A$ for the isotopic chain of Ni. The SLy4 (solid line), SGII
(dashed line), and Sk3 (dotted line) forces have been used in the
HF+BCS calculations of the densities in the case of Brueckner EDF.
\label{fig1}}
\end{figure}

\begin{figure}
\centering
\includegraphics[width=78mm]{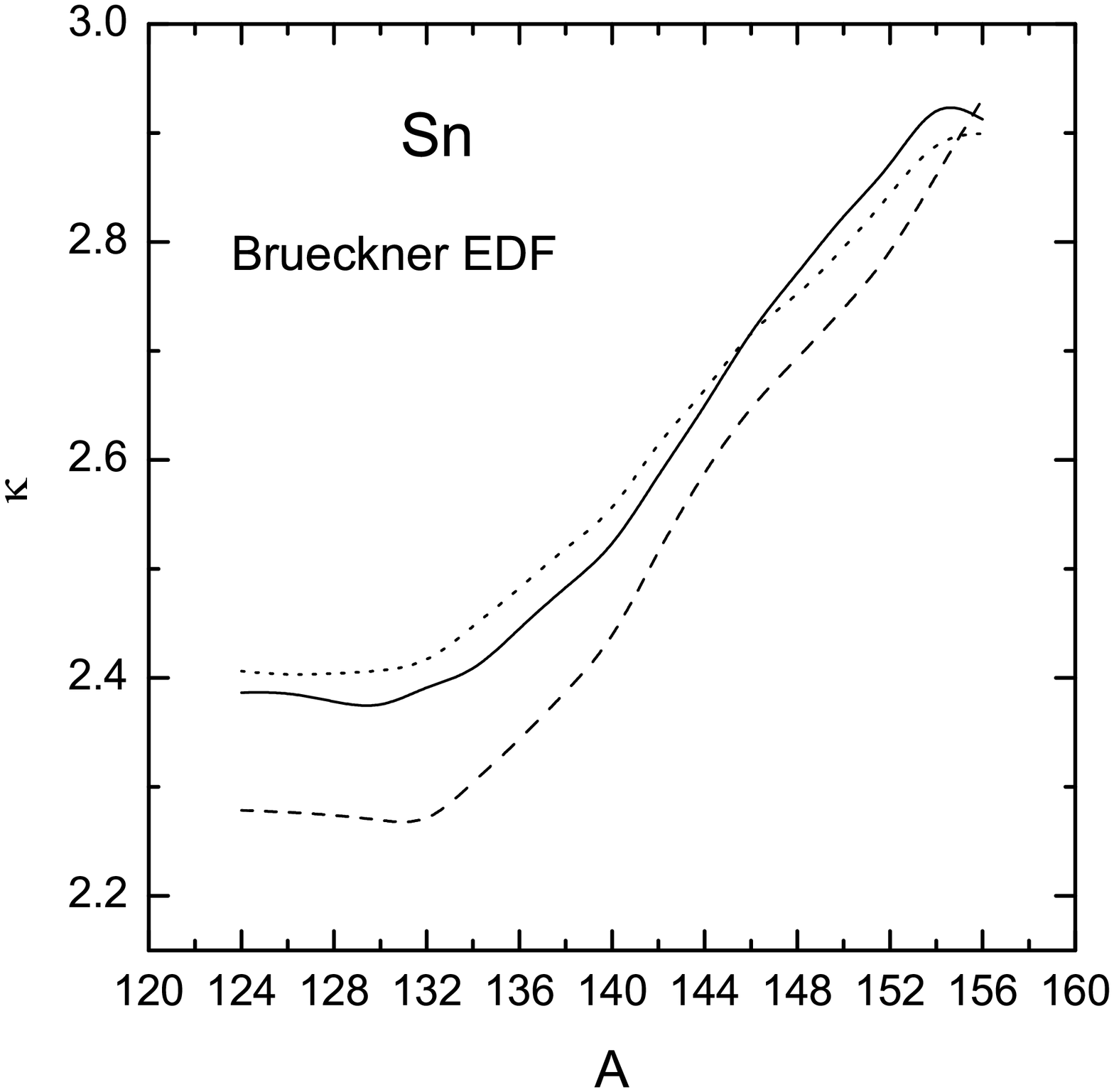}
\caption[]{The same as in Fig.~\ref{fig1} but for the isotopic
chain of Sn.
\label{fig2}}
\end{figure}

\begin{figure}
\centering
\includegraphics[width=78mm]{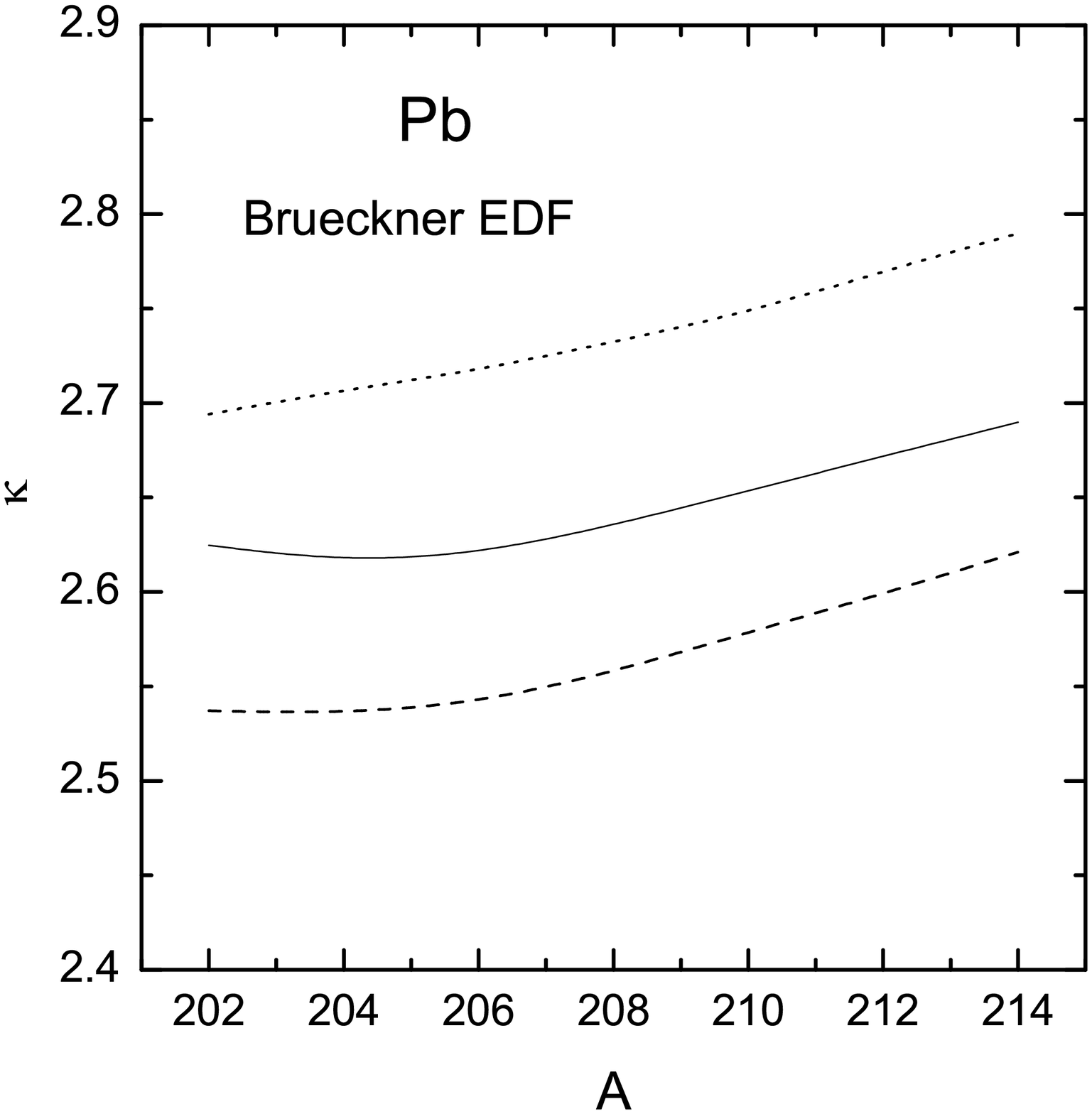}
\caption[]{The same as in Fig.~\ref{fig1} but for the isotopic
chain of Pb.
\label{fig3}}
\end{figure}

\begin{figure*}
\centering
\includegraphics[width=148mm]{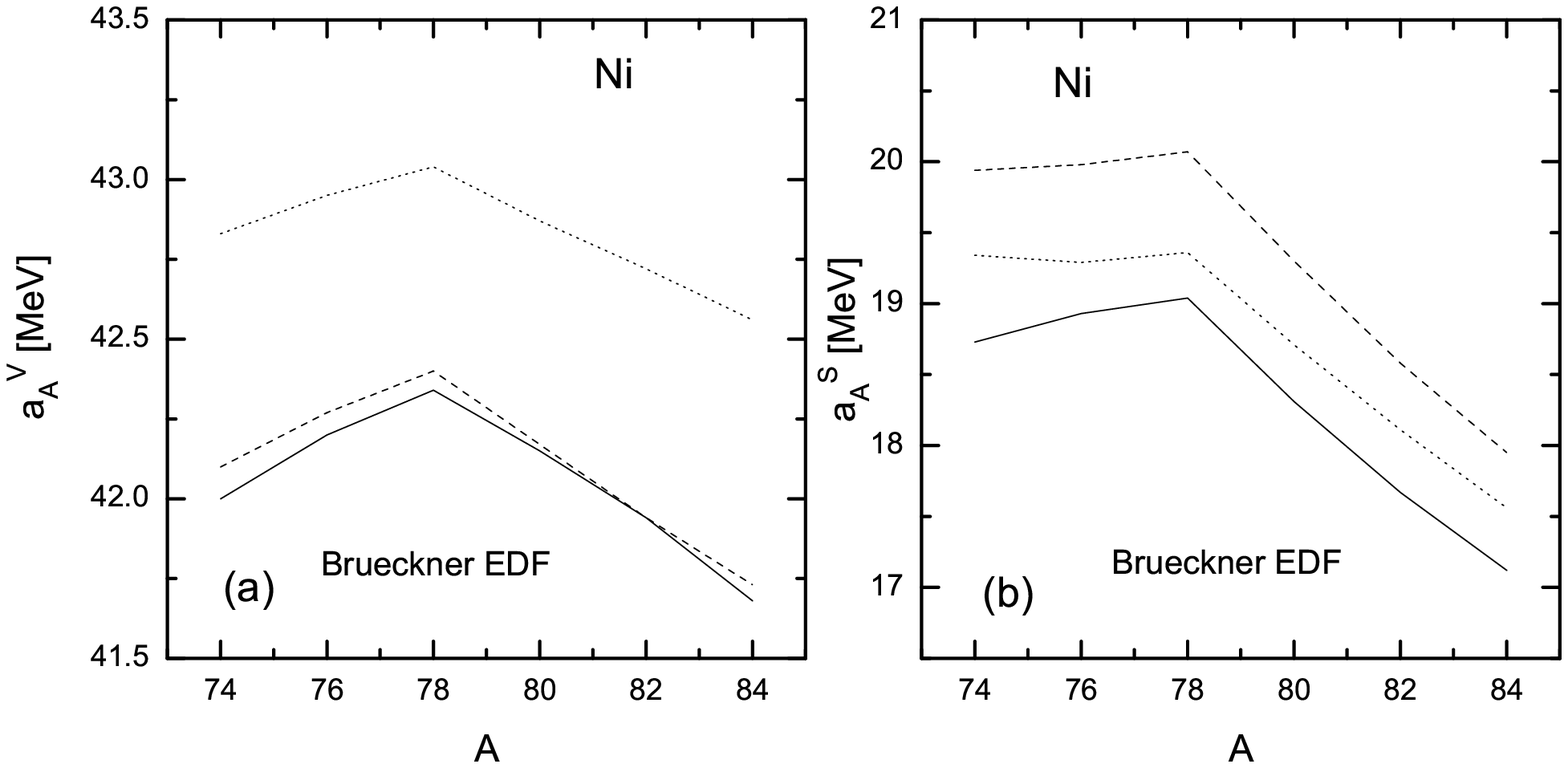}
\caption[]{The values of $a^{V}_{A}$ (a) and $a^{S}_{A}$ (b) as
functions of $A$ for the isotopic chain of Ni. The SLy4 (solid
line), SGII (dashed line), and Sk3 (dotted line) forces have been
used in the HF+BCS calculations of the densities in the case of
Brueckner EDF. \label{fig4}}
\end{figure*}

\begin{figure*}
\centering
\includegraphics[width=148mm]{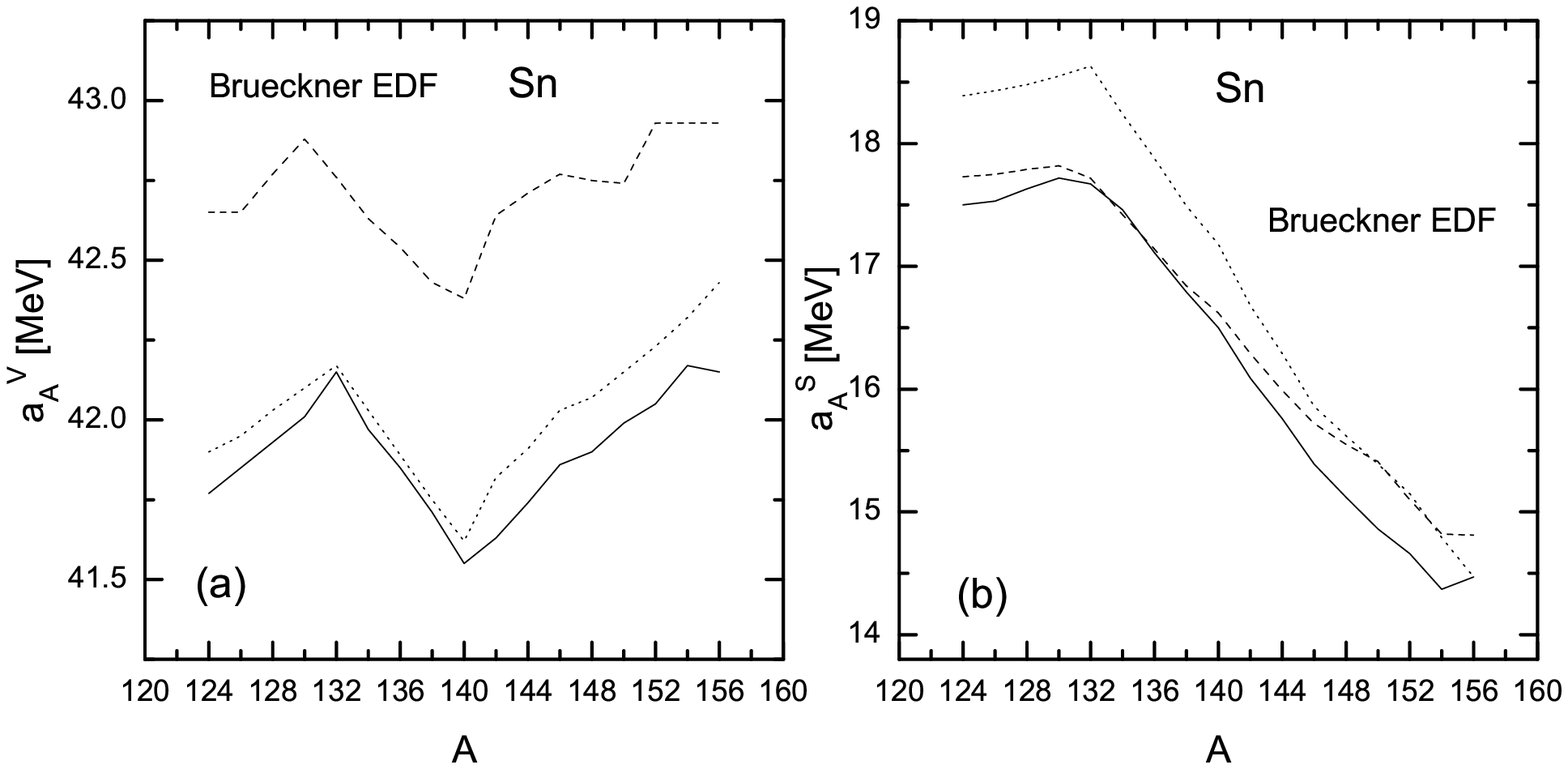}
\caption[]{The same as in Fig.~\ref{fig4} but for the isotopic
chain of Sn.
\label{fig5}}
\end{figure*}

\begin{figure*}
\centering
\includegraphics[width=148mm]{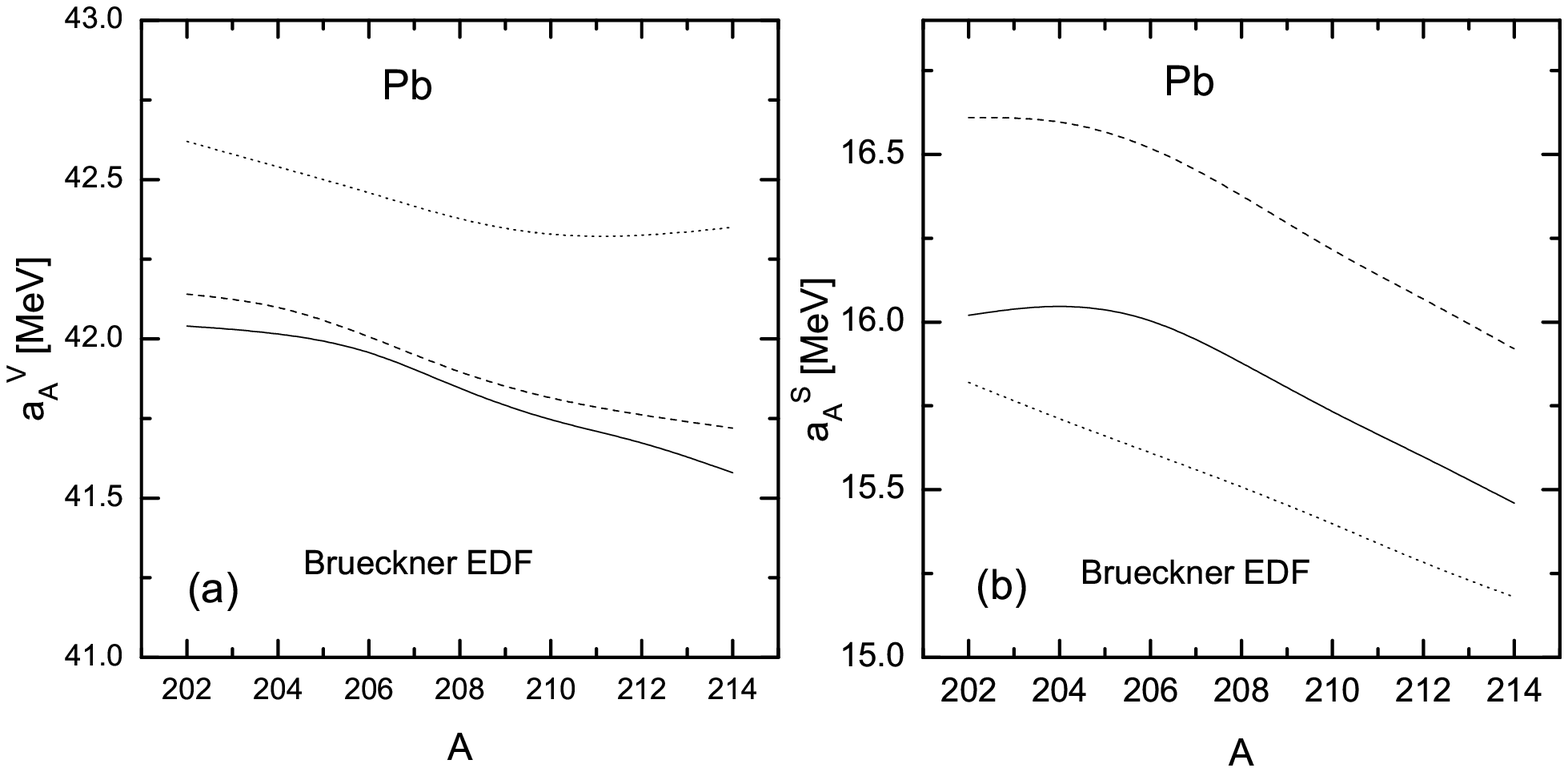}
\caption[]{The same as in Fig.~\ref{fig4} but for the isotopic
chain of Pb.
\label{fig6}}
\end{figure*}

It can be seen from Figs.~\ref{fig1}-\ref{fig3} that our results
for the values of the ratio $\kappa$ are within the range
\begin{equation}
2.10 \leq \kappa \leq 2.90 \;\;  .
\label{eq:29}
\end{equation}
This range of values is similar to the estimations of Danielewicz
{\it et al.} obtained from a wide range of available data on the
binding energies and from fits to other nuclear properties, such
as the neutron-skin thickness and the excitation energies to the
IAS \cite{Dan2004} (for definitions and examples of IAS, see,
e.g., Ref.~\cite{Tsang2012}). As already mentioned, it has been
shown in Ref.~\cite{Dan2003} that a combination of masses and
neutron-skin- sizes constrains the values of the volume symmetry
parameter between 27 and 31 MeV and the value of the volume to
surface-symmetry parameter ratio between 2.0 and 2.8. The minimum
value of the ratio obtained in Ref.~\cite{Dan2003} is 1.7. The
ranges of the published values of the ratio $\kappa$ extracted
from nuclear properties and presented in Ref.~\cite{Diep2007} are
(see Table II in \cite{Diep2007}):
\begin{equation}
2.6 \leq \kappa \leq 3.0
\label{eq:30}
\end{equation}
from IAS and skins \cite{Dan2004},
\begin{equation}
2.0 \leq \kappa \leq 2.8
\label{eq:30a}
\end{equation}
from masses and skins \cite{Dan2003}, and
\begin{equation}
1.6 \leq \kappa \leq 2.0
\label{eq:30b}
\end{equation}
from the analyses in Ref.~\cite{Diep2007} of masses and skins. As
can be seen the ranges (\ref{eq:30}) and (\ref{eq:30a}) are in a
good agreement with our results (see Eq.~(\ref{eq:29})).

As can be seen in Fig.~\ref{fig1} there exists a "kink" in the
curve of $\kappa\equiv a^{V}_{A}/a^{S}_{A}$ as a function of $A$
for the double-magic $^{78}$Ni nucleus. Such a "kink" exists also
for the double-magic $^{132}$Sn nucleus that can be seen in
Fig.~\ref{fig2}. Here we would like to note that the origin of the
"kinks" is in the different behavior of the density distributions
$\rho(r)$ for given isotopes. Namely the derivative of $\rho(r)$
determines the weight function $|{\cal F}(x)|^{2}$
[Eq.~(\ref{eq:18})] that takes part in the integrand of the
integral in Eq.~(\ref{eq:23}) giving the ratio $\kappa\equiv
a^{V}_{A}/a^{S}_{A}$. The peculiarities of $\rho(r)$ for the
closed shells lead to the existence of "kinks". As shown in more
details in Ref.~\cite{Gaidarov2012}, the same is the reason for
the "kinks" in the NSE $s(A)$, as can be seen in the expression
for it (see Eq.~(\ref{eq:19})). The "kink" of $s(A)$ at $^{78}$Ni
can be seen in Fig.~2 of Ref.~\cite{Gaidarov2011}, and the kink of
$s(A)$ for $^{132}$Sn in Fig.~8 of the same paper. In the case of
Pb isotopic chain (see Fig.~\ref{fig3}) such "kinks" do not exist
and this reflects the smooth behavior without "kinks" of $s(A)$
and related quantities for the Pb isotopic chain
\cite{Gaidarov2011,Gaidarov2012}.

It is seen from Figs.~\ref{fig4}-\ref{fig6} that our CDFM results
obtained with Brueckner EDF for $a^{V}_{A}$ are between 41.5 and
43 MeV, while for $a^{S}_{A}$ they are between 14 and 20 MeV.
These values are somewhat larger than those from other references
given above. As can be seen from Eqs.~(\ref{eq:27}) and
(\ref{eq:28}), these differences are due mainly to the somewhat
larger values of the NSE ($s$) for finite nuclei obtained within
the CDFM using the Brueckner functional, because our values for
the component $\kappa =a^{V}_{A}/a^{S}_{A}$ (that are between 2.1
and 2.9) are in the range obtained by other authors. The ranges of
changes of our results for $a^{V}_{A}$ and $a^{S}_{A}$ in the case
of the Brueckner EDF depend on the Skyrme forces used in the
HF+BCS calculations of the nuclear densities. They are given in
Table~\ref{table1} together with their corresponding average
values for each force and isotopic chain. These results can be
compared with the results obtained in:

\begin{table*}
\caption{The ranges of changes of $a_{A}^{V}$ and $a_{A}^{S}$ and
their average values for SLy4, SGII, and Sk3 forces used in the
HF+BCS calculations of the nuclear densities with Brueckner EDF
for the Ni, Sn, and Pb isotopic chains.}
{\begin{tabular}{ccccccccccccc} \hline \hline
Isotopic chain & & & NSE component & & & SLy4 & & & SGII & & & Sk3 \\
\hline
Ni  & & & $a_{A}^{V}$      & & & $41.7 \div 42.3$  & & & $41.7 \div 42.4$ & & & $42.6 \div 43$   \\
    & & & $a_{A}^{S}$      & & & $17.1 \div 19$    & & & $18 \div 20$     & & & $17.6 \div 19.4$ \\
    & & & $\bar a_{A}^{V}$ & & & 42.05             & & & 42.1             & & & 42.83            \\
    & & & $\bar a_{A}^{S}$ & & & 18.3              & & & 19               & & & 18.73            \\
\hline
Sn  & & & $a_{A}^{V}$      & & & $41.6 \div 42.2$  & & & $42.4 \div 43$   & & & $41.6 \div 42.4$ \\
    & & & $a_{A}^{S}$      & & & $14.4 \div 17.7$  & & & $14.8 \div 17.8$ & & & $14.5 \div 18.6$ \\
    & & & $\bar a_{A}^{V}$ & & & 41.9              & & & 42.71            & & & 42.02            \\
    & & & $\bar a_{A}^{S}$ & & & 16.27             & & & 16.5             & & & 16.92            \\
\hline
Pb  & & & $a_{A}^{V}$      & & & $41.6 \div 42$    & & & $41.7 \div 42.1$ & & & $42.3 \div 42.6$ \\
    & & & $a_{A}^{S}$      & & & $15.5 \div 16.1$  & & & $16 \div 16.6$   & & & $15.2 \div 15.8$ \\
    & & & $\bar a_{A}^{V}$ & & & 41.84             & & & 41.92            & & & 42.43            \\
    & & & $\bar a_{A}^{S}$ & & & 15.82             & & & 16.33            & & & 15.5             \\
\hline \hline
\end{tabular}}
\label{table1}
\end{table*}

Ref.~\cite{Dan2003}: $27 \leq \alpha = a^{V}_{A} \leq 31$ MeV;

Ref.~\cite{Dan2004}: $30.0 \leq a^{V}_{A} \leq 32.5$ MeV;

Ref.~\cite{Danielewicz}: $31.5 \leq a^{V}_{A} \leq 33.5$ MeV, $9
\leq a^{S}_{A} \leq 12$ MeV;

Ref.~\cite{Dan2014}: $30.2 \leq a^{V}_{A} \leq 33.7$ MeV.

\noindent It is shown in Ref.~\cite{Dan2014} that at $A\geq 30$
$a^{V}_{A} \approx 33.2$ MeV and $a^{S}_{A} \approx 10.7$ MeV.

For completeness and better comparison we list below also the
results presented in Ref.~\cite{Dan2014} making reference to
Ref.~\cite{Moller95} $a^{V}_{A}=39.73$ MeV and $a^{S}_{A}=8.48$
MeV, to Ref.~\cite{Groote76} $a^{V}_{A}=31.74$ MeV and
$a^{S}_{A}=11.27$ MeV, and to Ref.~\cite{Koura2005}
$a^{V}_{A}=35.51$ MeV and $a^{S}_{A}=9.89$ MeV. These results are
consistent with each other in the region $30 \leq A \leq 240$. In
the latter mass region the averaged values obtained are i)
$a^{V}_{A} \approx 35.34$ MeV and $a^{S}_{A} \approx 9.67$ MeV in
Ref.~\cite{Dan2014}; ii) $30 \leq a^{V}_{A} \leq 32.5$ MeV in
Ref.~\cite{Dan2006}; iii) $30 \leq a^{V}_{A} \leq 33$ MeV and
$a^{S}_{A} \approx 11.3$ MeV in Ref.~\cite{Dan2009}.

We would like to note that the same peculiarities (as for the
ratio $\kappa\equiv a^{V}_{A}/a^{S}_{A}$), namely "kinks", appear
in the cases of $a^{V}_{A}$ and $a^{S}_{A}$ as functions of the
mass number $A$. In Figs.~\ref{fig4}(a) and (b) one can see
"kinks" for $a^{V}_{A}$ and $a^{S}_{A}$, respectively, in the case
of the double-magic nucleus $^{78}$Ni. In Fig.~\ref{fig5}(a) a
"kink" appears for $a^{V}_{A}(A)$ not only for the double-magic
$^{132}$Sn, but also for the semi-magic $^{140}$Sn nucleus. The
latter is related to the closed $2f_{7/2}$ subshell for neutrons.
The behavior of $a^{S}_{A}(A)$ does not allow a corresponding
"kink" for $^{140}$Sn to be visible in the ratio $\kappa\equiv
a^{V}_{A}/a^{S}_{A}$. Reiss {\it et al.} discussed in
Ref.~\cite{Reis99} that the region around $N=90$ for neutron-rich
tin isotopes is an interesting one because the shell structure is
somewhat fluctuating. Although the average gap at $N=90$ was found
to be small, there are indications for a weak subshell closure,
the latter being supported also by the small jump in the
two-neutron separation energy in the same region, both calculated
at zero temperature \cite{Reis99}. In addition, in
Ref.~\cite{Nakada2014} $N=90$ was predicted to be submagic with
Gogny D1S and D1M interactions at $^{140}$Sn because the
$2f_{7/2}$ orbit is fully occupied, but not with M3Y-P6 and P7
semi-realistic NN interactions. As can be seen from
Eqs.~(\ref{eq:27}) and (\ref{eq:28}), the reason for "kinks" in
the separate coefficients as functions of $A$ is twofold. One of
them is the already mentioned reason for the "kinks" in the ratio
$\kappa\equiv a^{V}_{A}/a^{S}_{A}$, while the same reason causes
also "kinks" in the NSE ($s$) at closed-shell nuclei.

The second EDF that we use in the calculations is that one of
Skyrme with different Skyrme forces (e.g., Ref.~\cite{Skyrme}).
The aim to use this EDF is twofold: i) we can compare the values
of our $A$-dependent quantities $a^{V}_{A}$, $a^{S}_{A}$, and
$\kappa\equiv a^{V}_{A}/a^{S}_{A}$ obtained in the CDFM with the
$A$-independent ones obtained in the Danielewicz's approximation
(e.g., \cite{Dan2003,Danielewicz}) and, ii) the nuclear densities
are obtained using the same Skyrme forces in the HF+BCS
calculations, so there is a self-consistency of the approach. In
the standard Skyrme EDF the symmetry energy $s_{Sk}[\rho_{0}(x)]$
of the nuclear matter with density $\rho_{0}(x)$ can be expressed
by (e.g., Ref.~\cite{Wang2015}):
\begin{eqnarray}
s_{Sk}[\rho_{0}(x)]&=&\frac{\hbar^{2}}{6m}\left (\frac{3\pi^{2}}{2}\right )^{2/3} \rho_{0}^{2/3}(x)-\frac{1}{8}t_{0}(1+2x_{0})\rho_{0}(x) \nonumber \\
&-& \frac{1}{48}t_{3}(1+2x_{3})\rho_{0}^{\sigma+1}(x)-\frac{1}{24}\left (\frac{3\pi^{2}}{2}\right )^{2/3} \nonumber \\
&\times & [3t_{1}x_{1}-t_{2}(4+5x_{2})]\rho_{0}^{5/3}(x).
\label{eq:sskyrme}
\end{eqnarray}
The parameters of SGII, Sk3, and SLy4 Skyrme forces used to
calculate the symmetry energy of ANM, as well as the values of the
nuclear matter equilibrium density $\rho_{0}$, $r_{0}$, and the
symmetry energy at equilibrium density $s(\rho_{0})$, are listed
in Table~\ref{table2}. In addition, in the same Table we give the
values of the spin-orbit parameter $W_{0}$ used in the HF+BCS
calculations of the density distributions for the three Skyrme
forces.

\begin{table}
\caption{The parameters of SGII, Sk3, and SLy4 Skyrme forces in
the Skyrme EDF, the spin-orbit parameter $W_{0}$, the ANM
equilibrium density $\rho_{0}$, $r_{0}$, and the symmetry energy
at equilibrium density $s(\rho_{0})$.} {\begin{tabular}{cccccccc}
\hline \hline
& & & SGII & & Sk3 & & SLy4 \\
\hline
$t_{0}$ (MeV fm$^{3}$)         & & & -2645.0 & & -1128.75 & & -2488.91  \\
$t_{1}$ (MeV fm$^{5}$)         & & & 340.0   & & 395.0    & & 486.82    \\
$t_{2}$ (MeV fm$^{5}$)         & & & -41.9   & & -95.0    & & -546.39   \\
$t_{3}$ (MeV fm$^{3+3\sigma}$) & & & 15595.0 & & 14000.0  & & 13777.0   \\
$x_{0}$                        & & & 0.09    & & 0.45     & & 0.834     \\
$x_{1}$                        & & & -0.0588 & & 0        & & -0.344    \\
$x_{2}$                        & & & 1.425   & & 0        & & -1.0      \\
$x_{3}$                        & & & 0.06044 & & 1.0      & & 1.354     \\
$\sigma$                       & & & 0.16667 & & 1.0      & & 0.16667   \\
$W_{0}$                        & & & 105.0   & & 120.0    & & 123.0     \\
\hline
$\rho_{0}$ (fm$^{-3}$)         & & & 0.1583  & & 0.1453   & & 0.1595    \\
$r_{0}$ (fm)                   & & & 1.147   & & 1.18     & & 1.144     \\
$s(\rho_{0})$ (MeV)            & & & 26.84   & & 28.17    & & 32.01     \\
\hline \hline
\end{tabular}}
\label{table2}
\end{table}

The values of $\kappa$ for the three isotopic chains (of Ni, Sn
and Pb) obtained using SLy4, SGII, and Sk3 forces are given in
Figs.~\ref{fig7}, \ref{fig8}, and \ref{fig9} and those of
$a^{V}_{A}$ and $a^{S}_{A}$ are presented in Figs.~\ref{fig10},
\ref{fig11}, and \ref{fig12}. The ranges of changes of the latter,
as well as their average values, are given in Table~\ref{table3}.

\begin{figure}
\centering
\includegraphics[width=78mm]{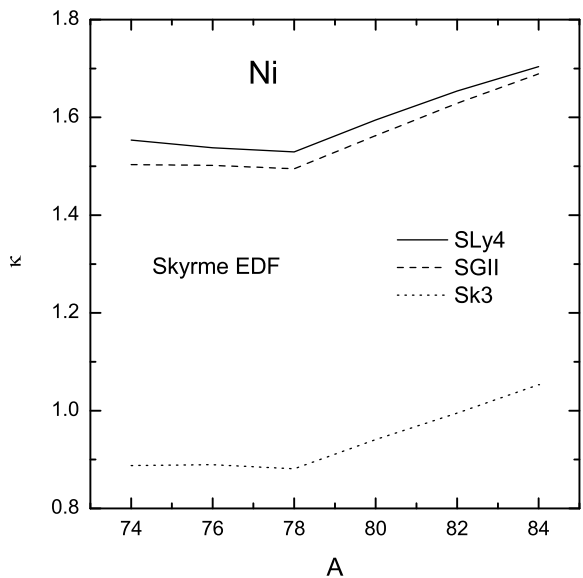}
\caption[]{The ratio $\kappa=a^{V}_{A}/a^{S}_{A}$ as a function of
$A$ for the isotopic chain of Ni in the case of Skyrme EDF with
use of SLy4, SGII, and Sk3 forces. \label{fig7}}
\end{figure}

\begin{figure}
\centering
\includegraphics[width=78mm]{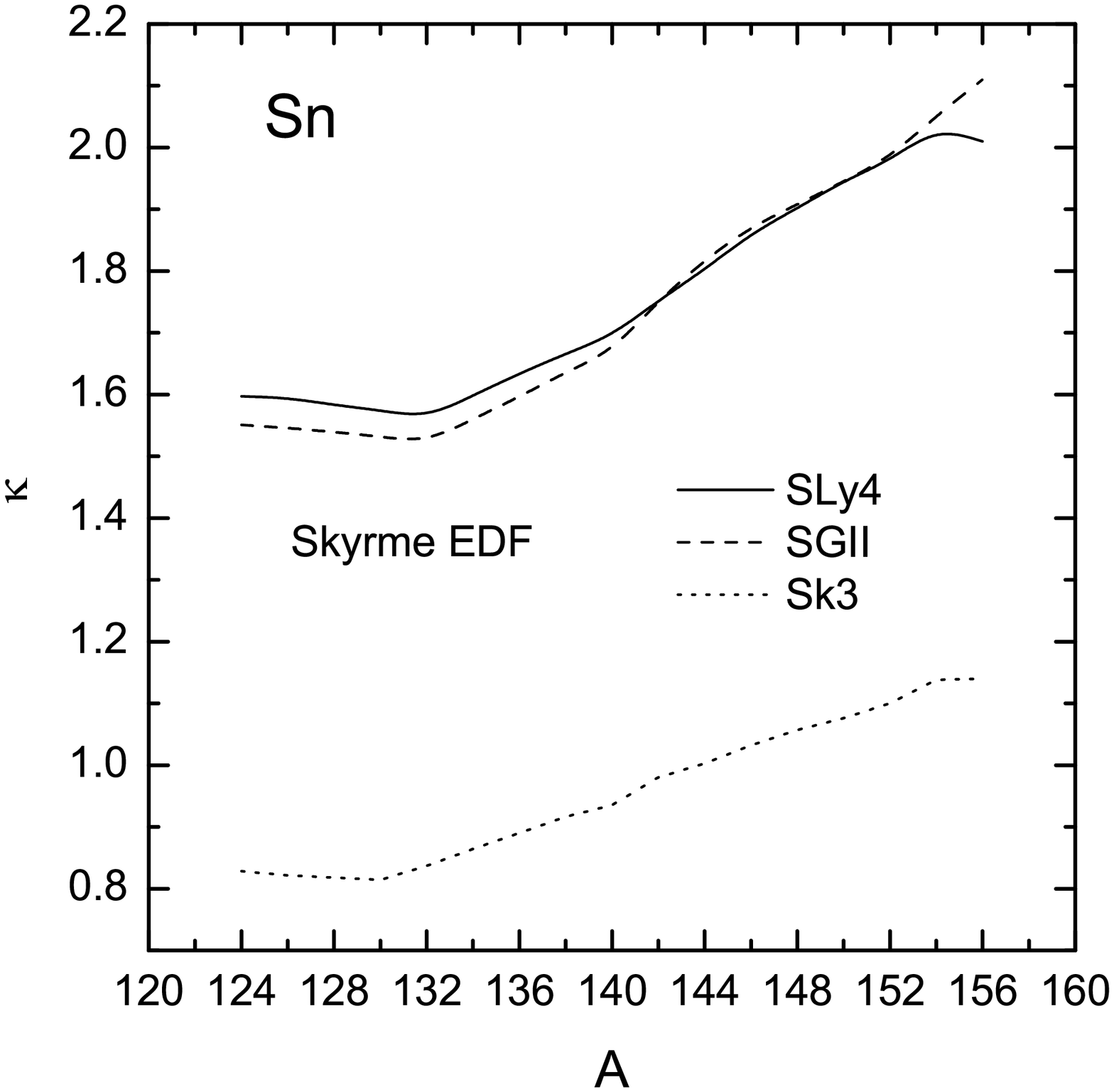}
\caption[]{The same as in Fig.~\ref{fig7} but for the isotopic chain of Sn.
\label{fig8}}
\end{figure}

\begin{figure}
\centering
\includegraphics[width=78mm]{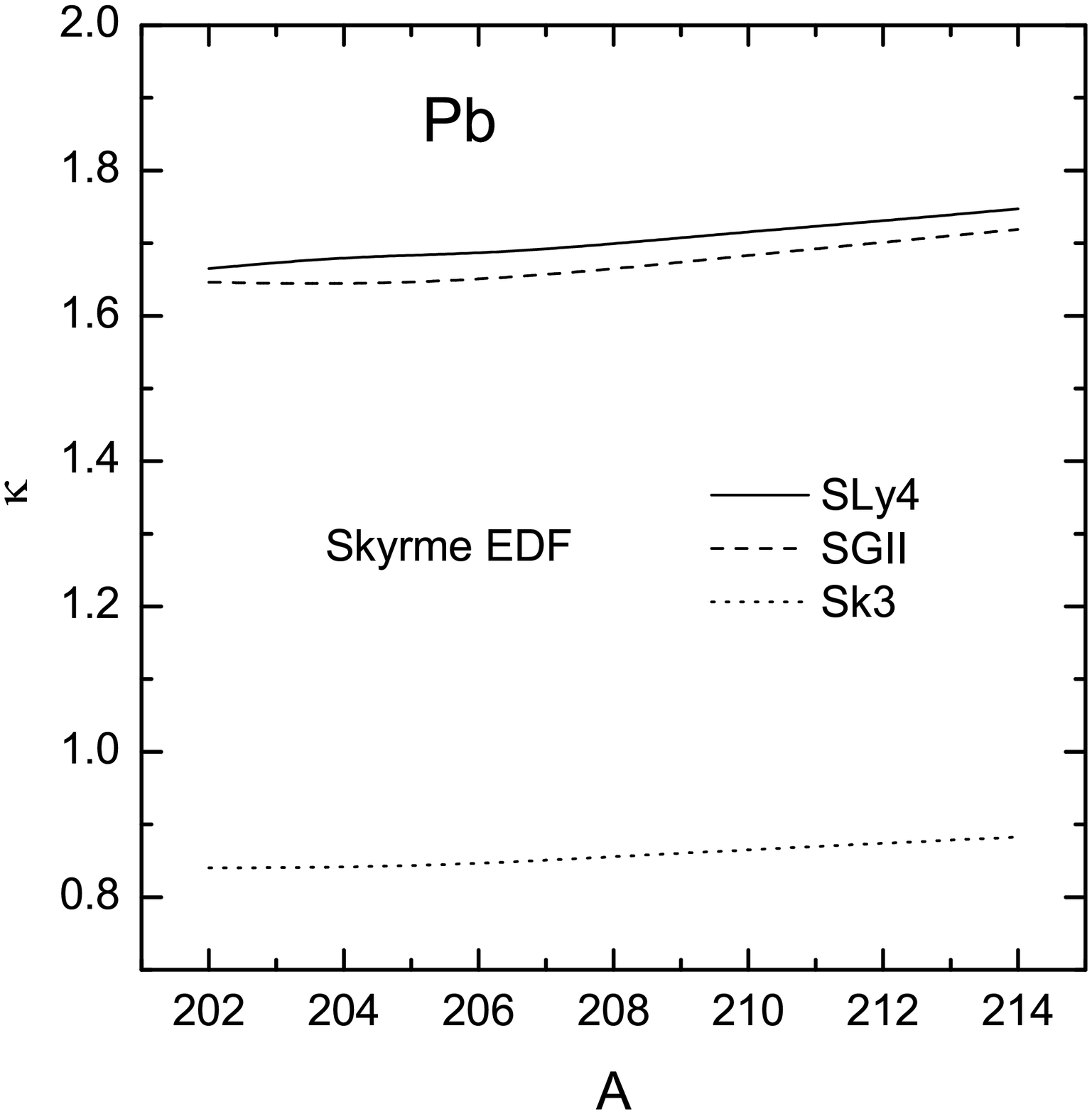}
\caption[]{The same as in Fig.~\ref{fig7} but for the isotopic chain of Pb.
\label{fig9}}
\end{figure}

\begin{figure*}
\centering
\includegraphics[width=148mm]{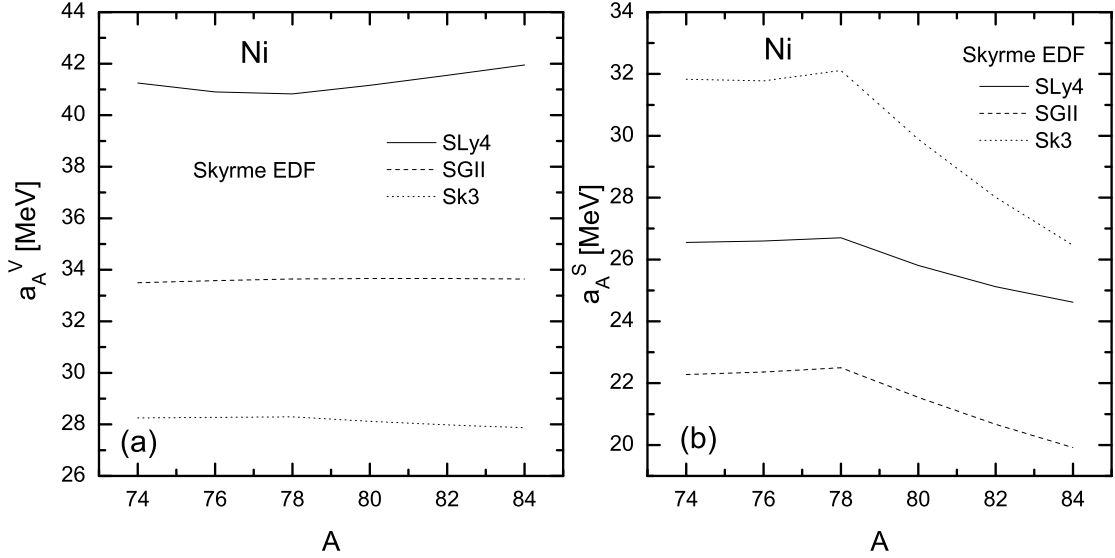}
\caption[]{The values of $a^{V}_{A}$ (a) and $a^{S}_{A}$ (b) as
functions of $A$ for the isotopic chain of Ni in the case of
Skyrme EDF with use of SLy4, SGII, and Sk3 forces.
\label{fig10}}
\end{figure*}

\begin{figure*}
\centering
\includegraphics[width=148mm]{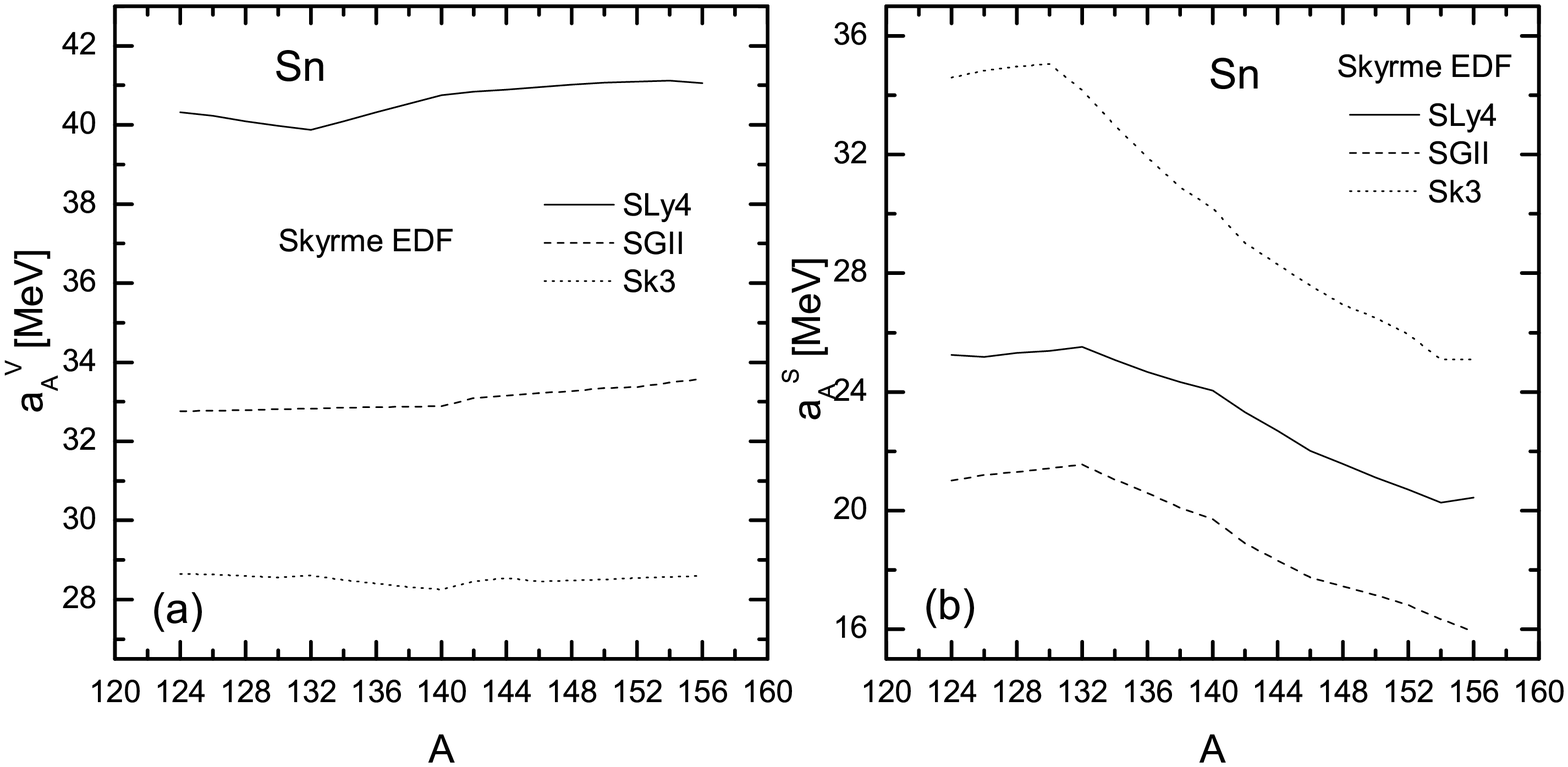}
\caption[]{The same as in Fig.~\ref{fig10} but for the isotopic chain of Sn.
\label{fig11}}
\end{figure*}

\begin{figure*}
\centering
\includegraphics[width=148mm]{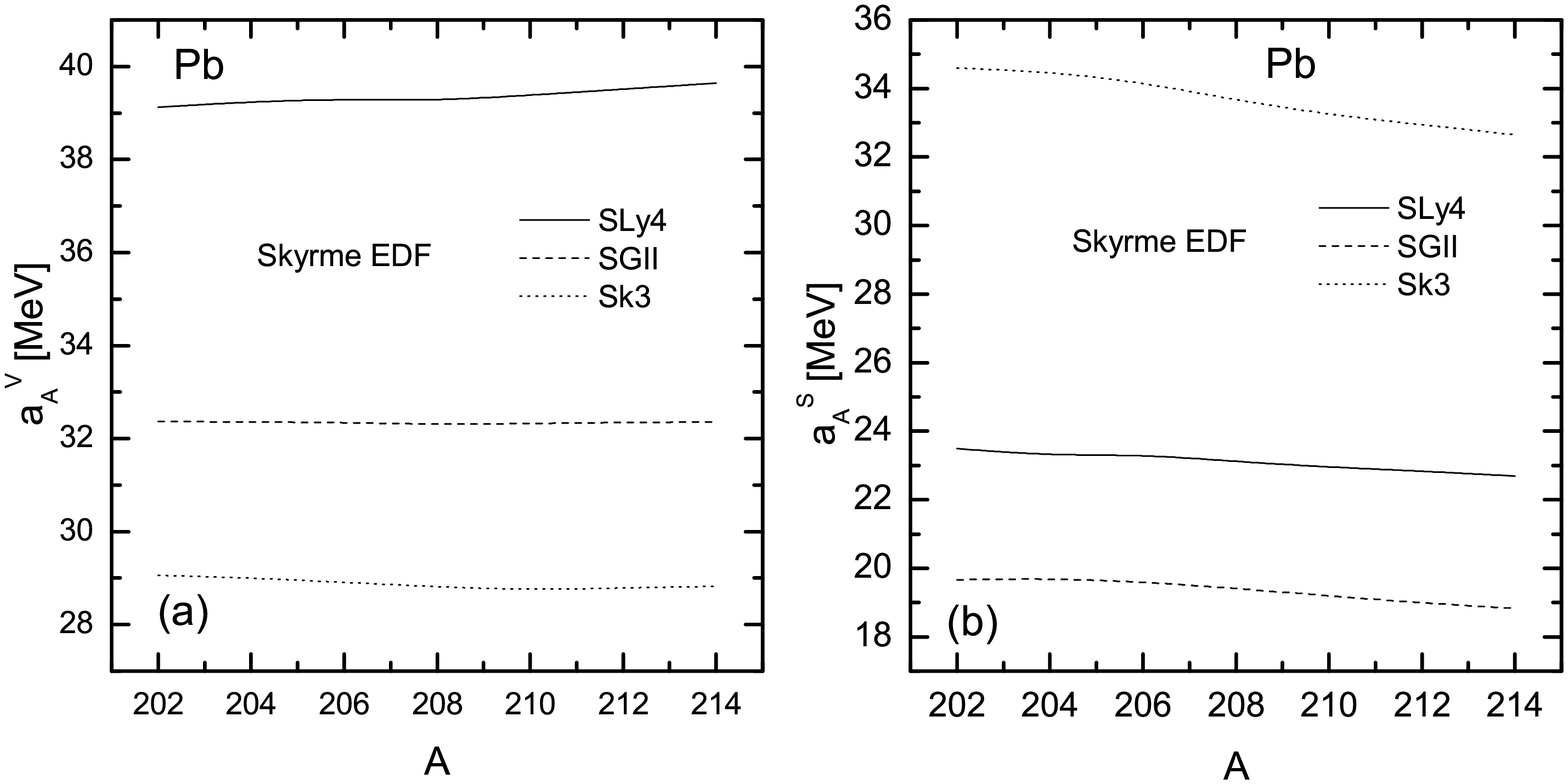}
\caption[]{The same as in Fig.~\ref{fig10} but for the isotopic chain of Pb.
\label{fig12}}
\end{figure*}

\begin{table*}
\caption{The ranges of changes of $a_{A}^{V}$ and $a_{A}^{S}$ and
their average values for SLy4, SGII, and Sk3 forces used in the
calculations with Skyrme EDF for the Ni, Sn, and Pb isotopic
chains.}
{\begin{tabular}{ccccccccccccc}
\hline \hline
Isotopic chain & & & NSE component & & & SLy4 & & & SGII & & & Sk3 \\
\hline
Ni  & & & $a_{A}^{V}$      & & & $40.8 \div 42$    & & & $33.5 \div 33.7$ & & & $27.9 \div 28.3$ \\
    & & & $a_{A}^{S}$      & & & $24.6 \div 26.7$  & & & $20 \div 22.5$   & & & $26.5 \div 32.1$ \\
    & & & $\bar a_{A}^{V}$ & & & 41.27             & & & 33.61            & & & 28.13            \\
    & & & $\bar a_{A}^{S}$ & & & 25.9              & & & 21.55            & & & 30.01            \\
\hline
Sn  & & & $a_{A}^{V}$      & & & $40 \div 41.1$    & & & $32.8 \div 33.6$ & & & $28.3 \div 28.6$ \\
    & & & $a_{A}^{S}$      & & & $20.3 \div 25.5$  & & & $16 \div 21.5$   & & & $25.1 \div 35$   \\
    & & & $\bar a_{A}^{V}$ & & & 40.6              & & & 33.06            & & & 28.51            \\
    & & & $\bar a_{A}^{S}$ & & & 23.36             & & & 19.21            & & & 30.24            \\
\hline
Pb  & & & $a_{A}^{V}$      & & & $39.1 \div 39.6$  & & & $32.3 \div 32.4$ & & & $28.8 \div 29.1$ \\
    & & & $a_{A}^{S}$      & & & $22.7 \div 23.5$  & & & $18.8 \div 19.7$ & & & $32.7 \div 34.6$ \\
    & & & $\bar a_{A}^{V}$ & & & 39.35             & & & 32.34            & & & 28.88            \\
    & & & $\bar a_{A}^{S}$ & & & 23.1              & & & 19.34            & & & 33.68            \\
\hline \hline
\end{tabular}}
\label{table3}
\end{table*}

As can be seen from Figs.~\ref{fig7}, \ref{fig8}, and \ref{fig9}
the ranges of changes of $\kappa$ are for the Ni isotopic chain:
$1.5 \leq \kappa \leq 1.7$ (SLy4 and SGII forces) and $0.88 \leq
\kappa \leq 1.05$ (Sk3 force), for the Sn isotopic chain: $1.52
\leq \kappa \leq 2.1$ (SLy4 and SGII forces) and $0.82 \leq \kappa
\leq 1.14$ (Sk3 force), and for the Pb isotopic chain: $1.65 \leq
\kappa \leq 1.75$ (SLy4 and SGII forces) and $0.84 \leq \kappa
\leq 0.88$ (Sk3 force). We note that the ranges of $\kappa$ for
the SLy4 and SGII forces in the three chains are in agreement with
those obtained in Ref.~\cite{Diep2007} $1.6\leq \kappa \leq 2.0$
from analyses of masses and skins.

What can be seen in Figs.~\ref{fig10}, \ref{fig11}, and
\ref{fig12} is that the values of $a^{V}_{A}$ are almost
independent on $A$ for a given isotopic chain and Skyrme force.
They are also similar in the different chains for a given Skyrme
force. The comparison of the results of our approach with those of
other authors shows that our values of $a^{V}_{A}$ for the
isotopic chains of Ni, Sn, and Pb for the SGII and Sk3 forces are
in agreement with those from, e.g.,
Refs.~\cite{Dan2003,Dan2014,Dan2009,Dan2006,Danielewicz} given
above, while the obtained values for the SLy4 force are comparable
with the results in Ref.~\cite{Moller95}. One can see also a
"kink" in the behavior of $\kappa$ for the Ni chain at $A=78$, for
the Sn chain at $A=132$ and a lack of "kinks" for the Pb chain,
like in the case when the Brueckner EDF is used. A "kink" at Ni
chain at $A=78$ can be seen also in the $A$-dependence of
$a^{S}_{A}$, as well as a "kink" of $a^{S}_{A}$ is seen at $A=132$
in the case of the Sn chain. In the latter small "kinks" can be
observed also for $a^{V}_{A}$ especially at $A=132$ for the SLy4
force. There are not "kinks" of $a^{V}_{A}$ and $a^{S}_{A}$ in the
Pb chain.

In the end of this section we would like to note that the obtained
values of $\kappa\equiv a^{V}_{A}/a^{S}_{A}$ in the CDFM
[Eq.~(\ref{eq:29})], that are in agreement with the recently
published values [Eq.~(\ref{eq:30})], are quite different from the
value of $\chi\equiv c_{4}/c_{3}$ from Eq.~(\ref{eq:3}) estimated
to be 1.1838 \cite{Myers66} or 1.14 \cite{Bethe71}. [We mention
that according to Eq.~(\ref{eq:22}) (and the text after it) for
large $A$: $c_{4}/c_{3}\simeq a^{V}_{A}/a^{S}_{A}$]. This
difference will be reflected in the corresponding values of
$c_{3}$ and $c_{4}$ that can be obtained using Eq.~(\ref{eq:8}).

\section{Conclusions}
\label{sec:conclusions}

In the present work we study the volume and surface components of
the NSE as well as their ratio within the framework of the CDFM.
This consideration is based on the calculations of the total NSE
that have been performed in our previous works
\cite{Gaidarov2011,Gaidarov2012,Gaidarov2014}, and also uses the
results of the earlier works on the subject (see, e.g.,
\cite{Myers66,Feenberg47,Cameron57,Bethe71,Green58}), as well as
the later theoretical approaches of Warda {\it et al.}
\cite{Warda2010}, Centelles {\it et al.} \cite{Centelles2010},
Danielewicz {\it et al.} (e.g.,
\cite{Dan2003,Dan2004,Dan2014,Dan2009,Dan2011,Tsang2012,Tsang2009,Ono2004,Dan2006,Danielewicz},
Dieperink and Van Isacker \cite{Diep2007} and others.

The results can be summarized as follows:

i) We develop, using as a base the Danielewicz's model
[Eq.~(\ref{eq:12})], another approach within the CDFM to calculate
the ratio $a^{V}_{A}/a^{S}_{A}$ between the volume and surface
components of the symmetry energy $s$, as well as $a^{V}_{A}$ and
$a^{S}_{A}$ separately, for finite nuclei. We obtain within the
CDFM the expression for $\kappa\equiv a^{V}_{A}/a^{S}_{A}$
[Eq.~(\ref{eq:23})] that allows us to calculate this ratio using
the ingredients of the model, the weight function $|{\cal
F}(x)|^{2}$ and the nuclear matter symmetry energy
$s^{ANM}[\rho_{0}(x)]$ from two energy- density functionals,
Brueckner and Skyrme ones. The first one of them was used to
calculate the NSE in our previous works
\cite{Gaidarov2011,Gaidarov2012,Gaidarov2014}. In the CDFM we take
nuclear matter values of the components of NSE to deduce their
values in finite nuclei. Thus, our approach is different from the
Danielewicz's formalism. Being motivated by the available
empirical data that show $A$-dependence of $a^{V}_{A}$,
$a^{S}_{A}$, and their ratio, we obtained in our approach a
possibility to find a (weak) $A$-dependence of the theoretical
results for these quantities within the CDFM. The weight function
$|{\cal F}(x)|^{2}$ [Eq.~(\ref{eq:18})] is calculated using the
proton and neutron density distributions obtained from the
self-consistent deformed HF+BCS method
\cite{Guerra91,Sarriguren2007} with density-dependent Skyrme
interactions;

ii) The values of $\kappa$ calculated using the Brueckner EDF for
the isotopic chains of Ni, Sn, and Pb are between 2.10 and 2.90.
This range of values is similar to the estimations of Danielewicz
{\it et al.} obtained from a wide range of available data on the
binding energies \cite{Dan2003}, of Steiner {\it et al.}
\cite{Steiner2005} and from a fit to other nuclear properties,
such as the excitation energies to IAS \cite{Dan2004},
neutron-skin thickness and others. The values of $\kappa$ obtained
using the Skyrme EDF for the same isotopic chains with SLy4, SGII,
and Sk3 forces are between 1.5 and 2.1 for the SLy4 and SGII
forces and between 0.82 and 1.14 for the Sk3 force. The former
result is in agreement with that obtained in Ref.~\cite{Diep2007}:
$1.6\leq \kappa \leq 2.0$ from the analyses of masses and skins;

iii) We calculate the values of the volume and surface
contributions to the NSE by means of Eqs.~(\ref{eq:27}) and
(\ref{eq:28}) within the CDFM. The values of NSE are taken from
our previous works \cite{Gaidarov2011,Gaidarov2012,Gaidarov2014},
where we used firstly the Brueckner EDF. The range of the values
obtained for the volume symmetry energy coefficient $a^{V}_{A}$
(between 41.5 and 43 MeV) is narrower than the one of the surface
symmetry energy coefficient $a^{S}_{A}$ (between 14 and 20 MeV).
The values of both coefficients are somewhat larger than the
already mentioned values of other works (see, e.g.,
\cite{Dan2003,Dan2004,Dan2014,Dan2009,Dan2006,Moller95,Groote76,Koura2005}
and others). We relate this difference to the larger values of the
total NSE for finite nuclei calculated by the use of the Brueckner
approach \cite{Brueckner68,Brueckner69} within the CDFM
\cite{Gaidarov2011,Gaidarov2012,Gaidarov2014}. Second, the
$a^{V}_{A}$ and $a^{S}_{A}$ are calculated within our approach
using the Skyrme EDF. It can be seen that the values of
$a^{V}_{A}$ are almost constant as functions of $A$ for a given
isotopic chain and Skyrme force. They are also similar in the
considered chains for a given Skyrme force. Our values of
$a^{V}_{A}$ for the isotopic chains of Ni, Sn, and Pb for the SGII
and Sk3 forces are in agreement with those from, e.g.,
Refs.~\cite{Dan2003,Dan2014,Dan2009,Dan2006,Danielewicz}, while
those in the case of the SLy4 force are similar to the results
presented in Refs.~\cite{Dan2014,Moller95}. We note that instead
of the Brueckner and Skyrme EDF's that are used in the present
work as examples, one can apply also other realistic functionals,
like the recently proposed Kohn-Sham EDF based on microscopic
nuclear and neutron matter equations of state \cite{Baldo2013};

iv) Studying firstly the isotopic sensitivity of $a^{V}_{A}$,
$a^{S}_{A}$, and their ratio in the case of using the Brueckner
EDF we observe peculiarities ("kinks") of these quantities as
functions of the mass number $A$ in the cases of the double-magic
$^{78}$Ni and $^{132}$Sn isotopes for $\kappa\equiv
a^{V}_{A}/a^{S}_{A}$, $a^{V}_{A}$, and $a^{S}_{A}$, as well as a
"kink" of $a^{V}_{A}$ for $^{140}$Sn. The latter is related to the
closed $2f_{7/2}$ subshell for neutrons. The origin of the "kinks"
is in the different behavior of the density distributions
$\rho(r)$ for the isotopes, because the derivative of $\rho(r)$
determines the weight function $|{\cal F}(x)|^{2}$
[Eq.~(\ref{eq:18})] that takes part in the expression for the
ratio $\kappa\equiv a^{V}_{A}/a^{S}_{A}$ [Eq.~(\ref{eq:23})]. As
shown in Ref.~\cite{Gaidarov2012}, the same is the reason for the
"kinks" in the NSE ($s$) [see Eq.~(\ref{eq:19})] observed in our
previous works \cite{Gaidarov2011,Gaidarov2012}. Similarly to the
case when Brueckner EDF is used, in the case of the Skyrme EDF one
can see also a "kink" in the behavior of $\kappa$ for the chain of
Ni at $A=78$ (Fig.~\ref{fig7}) and for the Sn chain at $A=132$
(Fig.~\ref{fig8}), as well as a lack of "kinks" for the Pb case
(Fig.~\ref{fig9}). A "kink" in the Ni chain at $A=78$ can be seen
also in the $A$-dependence of $a^{S}_{A}$ [Fig.~\ref{fig10}(b)],
as well as of $a^{S}_{A}$ at $A=132$ in the case of the Sn chain
[Fig.~\ref{fig11}(b)]. In Fig.~\ref{fig11}(a) small "kinks" can be
observed also for $a^{V}_{A}$ especially at $A=132$ for the SLy4
and Sk3 forces. "Kinks" of the $A$-dependence of $a^{V}_{A}$ and
$a^{S}_{A}$ in the Pb isotopic chain are not observed;

v) We show in subsection \ref{sec:cdfm} that, as expected, the
expression for the coefficient of the symmetry energy $a_{a}(A)$
[Eq.~(\ref{eq:10})] used, e.g. in
Refs.~\cite{Dan2003,Dan2004,Dan2006,Danielewicz,Diep2007}, can be
approximately written for large $A$ in the form of
Eqs.~(\ref{eq:2}), (\ref{eq:7}), and (\ref{eq:22}) introduced by
Cameron \cite{Cameron57} and used by Bethe \cite{Bethe71}, Myers
and Swiatecki \cite{Myers66} and others. We note that the obtained
values of $\kappa\equiv a^{V}_{A}/a^{S}_{A}$ in the CDFM using
Brueckner and Skyrme EDFs that are in agreement with the recently
published values are quite different from the value $\chi\equiv
c_{4}/c_{3}$ [Eq.~(\ref{eq:3})] that is estimated to be 1.1838
\cite{Myers66} or 1.14 \cite{Bethe71} (having in mind that for
large $A$ $c_{4}/c_{3}\simeq a^{V}_{A}/a^{S}_{A}\equiv \kappa$,
see Eq.~(\ref{eq:22})).

The suggested approach using the CDFM and based on a given EDF and
self-consistent mean-field method makes it possible to start with
the global values of parameters for infinite nuclear matter and to
derive their corresponding values in finite nuclei, which become
$A$-dependent. This is the main difference from other approaches.
The method uses the obtained within the CDFM symmetry energy
coefficient $s=a_{a}(A)$ in finite nuclei
\cite{Gaidarov2011,Gaidarov2012,Gaidarov2014} in the case of the
Brueckner EDF, as well as the NSE calculated in the present work
in the case of the Skyrme EDF. The calculation of the latter
avoids the problem related to fitting the Hartree-Fock energies to
LDM parametrization. The method makes it possible to obtain in the
present work additional information not only about the volume
contribution $a_{A}^{V}$ to the symmetry energy, but also about
the surface symmetry energy term $a_{A}^{S}$ of the LDM, as well
as to establish their eventual $A$-dependence. As known, the
$a_{A}^{S}$ is poorly constrained by empirical data. The obtained
results could provide a possibility to test the properties of the
nuclear energy density functionals and characteristics related to
NSE, e.g., the neutron skin thickness of finite nuclei.

\begin{acknowledgments}
Two of the authors (A.N.A. and M.K.G.) are grateful for support of
the Bulgarian Science Fund under Contract No.~DFNI-T02/19. E.M.G.
and P.S. acknowledge support from MINECO (Spain) under Contract
FIS2014-51971-P.
\end{acknowledgments}

\end{document}